ORIGINAL RESEARCH

# *Burkholderia* genome mining for nonribosomal peptide synthetases reveals a great potential for novel siderophores and lipopeptides synthesis

Qassim Esmaeel[1], Maude Pupin[2,3], Nam Phuong Kieu[4], Gabrielle Chataigné[1], Max Béchet[1], Jovana Deravel[1], François Krier[1], Monica Höfte[4], Philippe Jacques[1] & Valérie Leclère[1,2,3]

[1]University Lille, INRA, ISA, University Artois, University Littoral Côte d'Opale, EA 7394-ICV - Institut Charles Viollette, F-59000 Lille, France
[2]University Lille, CNRS, Centrale Lille, UMR 9189-CRIStAL, Centre de Recherche en Informatique Signal et Automatique de Lille, F-59000 Lille, France
[3]Bonsai Team, Inria-Lille Nord Europe, F-59655 Villeneuve d'Ascq Cedex, France
[4]Laboratory of Phytopathology, Faculty of Bioscience Engineering, Ghent University, Ghent, Belgium

**Correspondence**
Valérie Leclère, Institut Charles Viollette, ProBioGEM team, Bât. Polytech'lille, Avenue Langevin, Université de Lille1, F-59655 Villeneuve d'Ascq, Cedex, France.
Tel: +33320 43 46 68;
E-mail: valerie.leclere@univ-lille1.fr

**Funding Information**
The work was supported by the University of Lille 1, the INTERREG IV program France-Wallonie-Vlaanderen (Phytobio project), the bioinformatics platform Bilille, and Inria. QE received financial support from Sana'a University (Yemen).

Received: 17 December 2015; Revised: 26 January 2016; Accepted: 3 February 2016

*MicrobiologyOpen* 2016; 5(3): 512–526

doi: 10.1002/mbo3.347

## Abstract

*Burkholderia* is an important genus encompassing a variety of species, including pathogenic strains as well as strains that promote plant growth. We have carried out a global strategy, which combined two complementary approaches. The first one is genome guided with deep analysis of genome sequences and the second one is assay guided with experiments to support the predictions obtained in silico. This efficient screening for new secondary metabolites, performed on 48 gapless genomes of *Burkholderia* species, revealed a total of 161 clusters containing nonribosomal peptide synthetases (NRPSs), with the potential to synthesize at least 11 novel products. Most of them are siderophores or lipopeptides, two classes of products with potential application in biocontrol. The strategy led to the identification, for the first time, of the cluster for cepaciachelin biosynthesis in the genome of *Burkholderia ambifaria* AMMD and a cluster corresponding to a new malleobactin-like siderophore, called phymabactin, was identified in *Burkholderia phymatum* STM815 genome. In both cases, the siderophore was produced when the strain was grown in iron-limited conditions. Elsewhere, the cluster for the antifungal burkholdin was detected in the genome *of B. ambifaria* AMMD and also *Burkholderia* sp. KJ006. *Burkholderia pseudomallei* strains harbor the genetic potential to produce a novel lipopeptide called burkhomycin, containing a peptidyl moiety of 12 monomers. A mixture of lipopeptides produced by *Burkholderia rhizoxinica* lowered the surface tension of the supernatant from 70 to 27 mN·m$^{-1}$. The production of nonribosomal secondary metabolites seems related to the three phylogenetic groups obtained from 16S rRNA sequences. Moreover, the genome-mining approach gave new insights into the nonribosomal synthesis exemplified by the identification of dual C/E domains in lipopeptide NRPSs, up to now essentially found in *Pseudomonas* strains.

## Introduction

The *Burkholderia* genus includes more than 60 species that colonize a wide range of environments including soil, water, plants, animal, and human (Mahenthiralingam et al. 2008). They are well known for their nutritional versatility, which certainly contributes to their capacity to live in extreme and diverse habitats and which has led to their use in biocontrol, bioremediation processes, and biodegradation of pollutants (Mahenthiralingam and Goldberg 2005). Some *Burkholderia* species are involved in promoting plant or nitrogen-fixing bacteria (Caballero-Mellado et al. 2007). On the other hand, several species of *Burkholderia*, including those from *B. cepacia* complex group (Bcc) as well as *B. gladioli* and *B. fungorum*, represent a significant threat to the life of immunocompromised individuals, especially for those who are suffering from cystic fibrosis or chronic granulomatous diseases (Coenye et al. 2001). *Burkholderia pseudomallei* and *Burkholderia mallei* are the only known members of the genus *Burkholderia* that are primary pathogens in humans and animals, causing melioidosis in humans (Cheng and Currie 2005) and glanders in horses (Nierman et al. 2004). The ability to adapt and colonize a wide variety of environments is likely due to an unusually large, complex, and variable genome (4.6–9 Mb), split into up to three chromosomes and large plasmids. Furthermore, some species have been found to secrete a variety of extracellular enzymes with proteolytic, lipolytic, and hemolytic activities (Vial et al. 2007), together with secondary metabolites. These include siderophores called ornibactins and malleobactins (Franke et al. 2014), hemolytic peptides called cepalysins (Abe and Nakazawa 1994), antifungal agents as pyrrolnitrin (Hammer et al. 1999), cepapafungins, and related compounds cepacidines A1 and A2 (Lim et al. 1994). More recently, antifungal cyclic lipopeptides (CLPs) called occidiofungins/burkholdines (bks) have been isolated from strains belonging to the Bcc group (Gu et al. 2009; Tawfik et al. 2010). Some secondary metabolites including siderophores and peptides can be produced nonribosomally by NonRibosomal Peptide Synthetases called NRPS (Marahiel 2009). These large synthetases are organized into sets of domains that constitute modules containing the information needed to complete an elongation step in an original peptide biosynthesis. The main catalytic functions are responsible for the activation of an amino acid residue (adenylation, -A domain), the transfer of the corresponding adenylate to the enzyme-bound 4-phosphopantetheinyl cofactor (thiolation, -T domain), the peptide bond formation (condensation, -C domain), and the release of the peptide from the NRPS (thioesterase, -Te domain). Optional domains of NRPS modules can modify the amino acids by different reactions such as *N*-methylation, epimerization (-E domain), or heterocyclization.

A genome mining was conducted with the aim of identifying biosynthetic clusters for secondary metabolites, but in this study, we have chosen to only present the clusters including at least one gene encoding NRPS. The domain organization of each NRPS protein was designed, the C-domain sub-types were determined, and the prediction of the most probable produced peptides was performed in silico using specific bioinformatics tools included in the Florine workflow previously described (Caradec et al. 2014). We specially describe the identification of siderophores and lipopeptides because of their possible use in biocontrol (Sharma and Johri 2003; Roongsawang et al. 2011). Many of the structure and activity predictions obtained in silico are corroborated by experimental supports as biological activities from wild-type strains and deletion mutants, iron chelating measurements, surface tension, or MS detections of the compounds in growth supernatants.

## Materials and Methods

### In silico detection of NRPS genes within sequenced genomes

Forty-eight gapless complete genomes of *Burkholderia* strains available in the NCBI were mined by following the strategy described within the Florine workflow (Caradec et al. 2014). Sequences presenting identity with known NRPSs were detected by using the NCBI-Blastp tool with the *B. cepacia* GG4 OrbJ used as a query (YP_006615894). The automatically annotated proteins were also fished using a list of keywords including adenylation, synthetase, thiotemplate, phosphopantetheine, nonribosomal, NRPS, siderophores, ornibactin, pyochelin, syringomycin, malleobactin, polyketide, and acyl carrier protein (ACP). Identification and annotation of NRPS gene clusters in all selected strains were also detected by antiSMASH (http://antismash.secondarymetabolites.org) (Weber et al. 2015).

Protein sequences of all genes encoding NRPSs were obtained from NCBI databases. The modular organization and domain architecture were deciphered combining the results from NRPS-PKS analysis website (http://nrps.igs.umaryland.edu/nrps) (Bachmann and Ravel 2009) and Structure-Based Sequence Analysis of Polyketide Synthases (SBSPKS) (http://www.nii.ac.in/~pksdb/sbspks/search_main_pks_nrps.html) (Anand et al. 2010) with antiSMASH results. The specificity of the A domain was conducted by using the web-based software NRPSpredictor2 (http://nrps.informatik.uni-tuebingen.de) (Röttig et al. 2011). The C-domain types were determined by identification of specific signatures of the DownSeq (Caradec et al. 2014). The D-configuration of monomers of nonribosomal peptide was predicted regarding the presence of E followed by $^D C_L$ domains or dual C/E- domains.

The structures of the predicted peptides were compared to other nonribosomal peptides through the structural search tool of the Norine database (bioinfo.lifl.fr/norine) (Caboche et al. 2008, 2009).

### Culture conditions and media

Strains, plasmids, and primers used in this study are listed in Table S1. Molds (*Fusarium oxysporum*, *Galactomyces geotrichum*, *Botrytis cinerea*, and *Rhizoctonia solani*) were gown on PDA (Sigma-Aldrich, St Louis, MO). *Burkholderia* strains, *Micrococcus luteus*, *Listeria innocua*, and *Escherichia coli* were routinely grown at 37ºC on Luria Bertani (LB) medium. *Candida albicans* was grown on Sabouraud agar at 37ºC and *Saccharomyces cerevisiae* was grown on

yeast-extract–peptone–dextrose at 30°C. For lipopeptide production, strains were grown in Landy medium buffered with MOPS 100 mmol/L (Landy et al. 1948). When required, gentamycin sulfate (Sigma Aldrich) was added at a concentration of 25 $\mu$g mL$^{-1}$ for *E. coli* and at 300 $\mu$g mL$^{-1}$ for *B. ambifaria* AMMD *wt* and mutant strains.

## Construction of AMMD mutant strains

*Burkholderia ambifaria* AMMD-Δbamb_6472 was constructed using in vivo homologous recombination in the yeast, *S. cerevisiae* InvSc1 (Shanks et al. 2006). To construct *B. ambifaria* AMMD-Δbamb_6472, two regions of the NRPS gene Bamb_6472 were amplified using primers Up6472-F and Up6472-R (1015 bp amplicon) and primers Down6472-F and Down6472-R (887 bp amplicon), respectively. Deletion of NRPS gene Bamb_6472 was confirmed by PCR (See Protocole S2 for details).

## Biological activity assays

### Antifungal activity

Antifungal activity was demonstrated by contact antagonism assay. PDA plates were inoculated with 5 mm of agar plug of *F. oxysporum*, *G. geotrichum*, *B. cinerea*, and *R. solani* grown for 2 days on PDA. The plates were inoculated with the bacterium and incubated at 25°C for 5–7 days.

### Antiyeast activity

Antiyeast activity was performed by the agar drop plate method. Erlenmeyer flask containing 100 mL of PDB were inoculated with 1.5 mL of bacterial suspension containing about 10$^8$ CFU mL$^{-1}$ and incubated under rotary shaking (160 rpm) at 37°C for 5 days. Then, the culture was centrifuged at 10,000$g$ for 10 min at 4°C. The supernatant was concentrated 10 times via speed vacuum, and then 50 $\mu$L of this supernatant were deposited on 14 mL solid Sabouraud medium inoculated with target strains including *C. albicans* and *S. cerevisiae*. After 24 h of incubation at 30°C, the activity was estimated by measuring the diameter of the zone of inhibition of target strains.

### Hemolytic activity

The hemolytic activity was assessed at 30°C on LB supplemented with 5% (vol/vol) horse blood. Supernatant (50 $\mu$L) were spotted on the plates that were incubated for 48 h at 30°C. Positive results were detected by the formation of a clear zone around the wells.

## MALDI-ToF mass spectrometry

Bacterial sample were cultured in liquid medium and incubated at 37°C for 5 days. Supernatant was mixed with a matrix solution (10 mg/mL cyano-4-hydroxycinnamic acid in 70% water, 30% acetonitrile, and 0.1% TFA). The samples were homogenized on a Vortex and centrifuged at 4500 g. For classical analysis 1 $\mu$L of sample solution was spotted onto a MALDI-ToF MTP 384 target plate (Bruker Daltonik GmbH, Leipzig, Germany) according to the procedure of the dried-droplet preparation. Mass profiles experiments were analyzed with an Ultraflex MALDI-ToF/ToF mass spectrometer (Bruker, Bremen, Germany) equipped with a smartbeam laser. Samples were analyzed using an accelerating voltage of 25 kV and matrix suppression in deflexion mode at m/z 750. The laser power was set to just above the threshold of ionization (around 35%). Spectra were acquired in reflector positive mode in the range of 400 at 3000 Da. Each spectrum was the result of 2000 laser shots per m/z segment per sample delivered in 10 sets of 50 shots distributed in three different locations on the surface of the matrix spot. The instrument was externally calibrated in positive reflector mode using bradykinin [M+H]+ 757.3991, angiotensin II [M+H]+ 1046.5418, angiotensin I [M+H]+ 1296.6848, substance P [M+H]+ 1347.7354, bombesin [M+H]+ 1619.8223, and ACTH (1-17) [M+H]+ 2093.0862.

## Production of siderophores

Strains were propagated three times in iron-deficient minimum medium (MM9) (Payne 1994) supplemented with 10% of casamino acids previously treated, first with 3% 8-hydroxyquinoline in chloroform to remove contaminating iron and then with chloroform to remove remaining 8-hydroxyquinoline. Then, MgCl$_2$ 1 mmol/L and CaCl$_2$ 0.1 mmol/L were added. All glassware was rinsed with HCl 6 mol/L. CAS (Chrome Azurol S) liquid and agar assays were performed in accordance to the original protocol (Schwyn and Neilands 1987).

## Surface tension measurements

The surface tension was measured according to the De Nouy methodology using a tensiometer TD1 (Lauda, Königshofen, Germany) as previously described (Leclère et al. 2006).

## Phylogeny tree

The 16S rRNA sequences were extracted from RDP database (Cole et al. 2009) and submitted to MEGA6 program (Tamura et al. 2013). The tree was built using the neighbor-joining method (Saitou and Nei 1987). The evolutionary distances were computed using the maximum composite likelihood method (Tamura et al. 2004). The rate variation among sites was modeled with a gamma distribution with five rate categories. Bootstrap analysis with 1000 replicates was performed to assess the support of the clusters (Felsenstein 1985). All positions containing gaps and missing data were eliminated. *Cupriavidus taiwanensis* was used as an outgroup reference.

## Results

### Overview on *Burkholderia* sequenced genomes

Forty-eight gapless complete genomes of *Burkholderia* strains available in 2014 within the NCBI genome bank were mined for NRPSs. To evaluate the relevancy of sequenced strains among the *Burkholderia* genus, the 16S rRNA sequences were extracted and aligned to build a phylogenetic tree of all sequenced strains (Fig. 1). The considered strains for NRPS seeking are representative of the genus as they include species playing ecological role related to xenobiosis, biocontrol, promotion of plant growth, biological nitrogen fixation, as well as pathogenic strains for animals as some of those belonging to the Bcc group. Elsewhere, the tree clearly shows that the strains are distributed into three clades. The first clade (clade I) includes *B. mallei*, *B. pseudomallei*, and the rarely pathogenic *B. thailandensis* strains, the second clade (clade II) gathers the Bcc group strains. The remaining species are contained in the clade III.

All the strains have their genome split into up to three chromosomes added with one or two plasmids, exceptionally three (*B.* sp.YI23), four (*B. gladioli* BSR3 and *B. glumae* BGR1), or five (*B. vietnamiensis* G4). The whole genome sizes span from 3.75 to 9.73 Mb with most frequent values between 7 and 9 Mb (33 strains).

Except for *B. phenoliruptrix* BR3459a and *B.* sp. CCGE1002, genes for at least one NRPS could have been detected in the genomic sequence of the 48 studied strains, whatever their lifestyle or virulence ability (Table 1).

### Identification of a novel malleobactin-like siderophore

A cluster of genes including those for NRPS responsible for ornibactin or malleobactin synthesis has been found in the genome of 41 strains (Table 1) (Fig. 2A). The cluster is present on the larger chromosome (chromosome 1), except for *B. phytofirmans* PsJN and *B. xenovorans* LB400 where it is located on chromosome 2, and for *B.* sp. YI23 and *B.* sp. RPE64 harboring the cluster on a plasmid. A complete cluster includes genes responsible for siderophore biosynthesis (coding for 2 NRPSs and accessory enzymes modifying the monomers of the peptidic chain), genes implicated in the uptake (siderophore export and ferri-siderophore TonB dependent receptor) and the *orbS* or *mbaF* genes coding for an extracytoplasmic sigma70 factor implicated in the regulation (Table S2). The NRPSs responsible for the siderophore backbone are encoded by *orbI-orbJ* and *mbaA-mbaB* in the case of ornibactin and malleobactin synthesis, respectively. The proteins harbor the same domain organization including 4 adenylation domains (A1-A4). The specificity code for the A1 domain is variable: Leu for ornibactin and Bht (beta-hydroxy-tyrosin) for malleobactin (Fig. 2A). According to the presence of an acyl transferase gene (*orbL*) together with the Leu selectivity of A1 domain, ornibactin is predicted to be produced by all the strains belonging to the Bcc group (*B. ambifaria*, *B. cenocepia*, *B. cepacia*, *B. lata*, *B. multivorans*, *B. vietnamiensis* and *B.* sp. KJ006) and two strains belonging to the clade III (YI23 and RPE64). The malleobactin synthetic cluster detected with the presence of an accessory enzyme gene (*mbaM*) and the Bht selectivity for A1 is found in the genomes of two strains belonging to the clade III (*B. phytofirmans* PsJN and *B. xenovorans* LB400), and the strains from clade I, except *B. thailandensis* 2002721723.

*Burkholderia phymatum* STM815 belonging to clade III, possesses on chromosome 2, a cluster with some characteristics for malleobactin synthesis and uptake (absence of the acyl transferase gene *orbL*), together with features related to ornibactin biosynthesis (absence of *mbaM* and presence of *orbK*). Moreover, the predicted specificities of A1 and A4 domains of the NRPS are Asp and Cys, respectively. These predictions are different from those obtained both with ornibactin and malleobactin NRPS A1 and A4 domains (Fig. 2A). The strain *B. phymatum* STM815 was grown under iron-limited conditions and was analyzed for siderophore production. The CAS agar assay was positive showing an orange halo surrounding the strain (Fig. S1) and compounds with a mass comprised between 445 Da and 773 Da were detected and fragmented from supernatant. No fragmentation spectrum could have been assigned to a known siderophore belonging to ornibactin or malleobactin family. We suggest to name this probably new malleobactin-like siderophore phymabactin and the corresponding *nrps* genes *phmA* and *phmB* (Fig. 2) (Table S2).

### Identification of a cluster involved in cepaciachelin synthesis

Pyochelin synthetase is constituted of two NRPSs of about 1360 and 1890 AA, harboring a specific domain architecture that includes two cyclization domains (Cy), associated

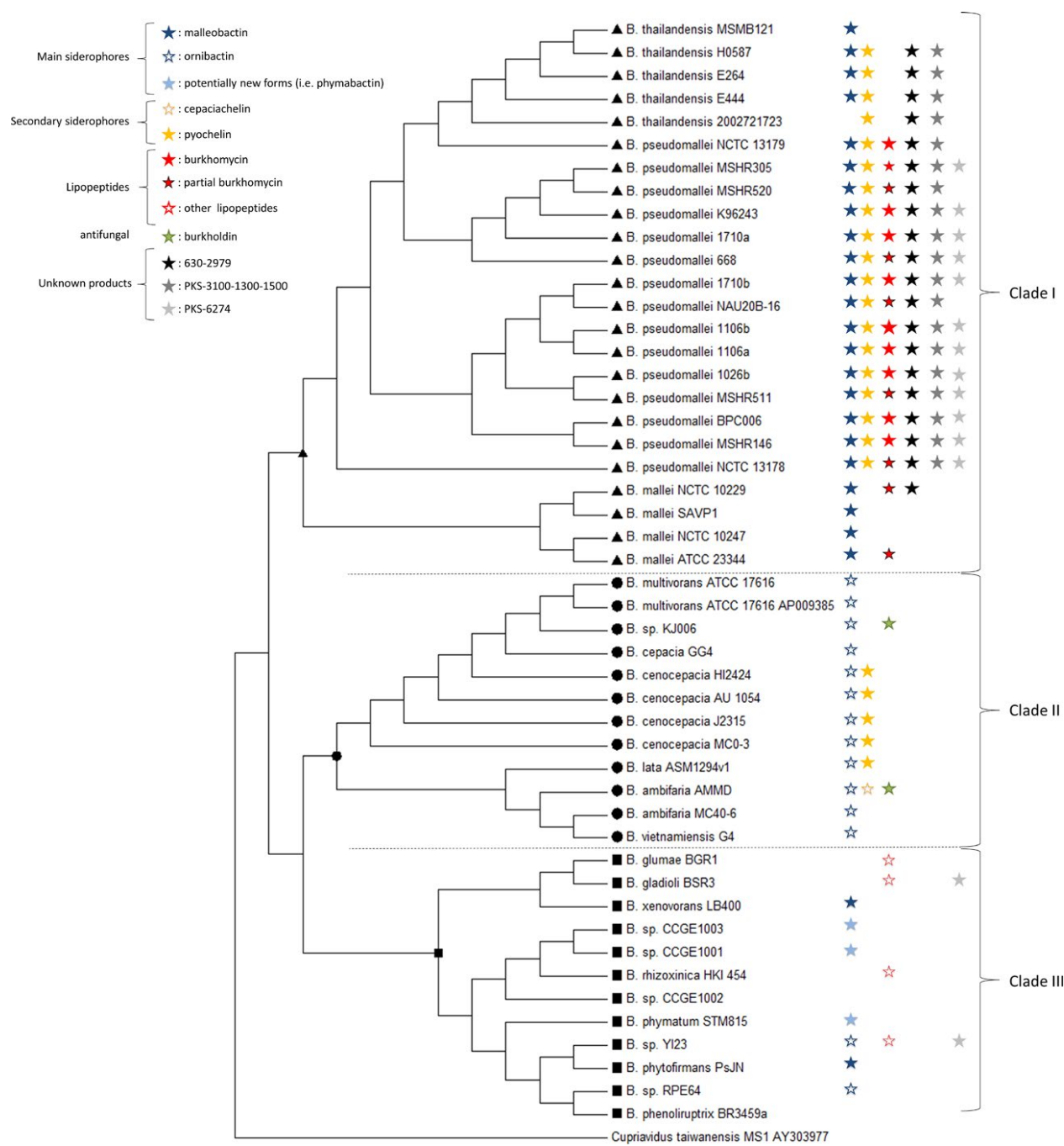

**Figure 1.** Phylogenetic tree of the *Bukholderia* strains with complete genome sequenced. Performed using MEGA6 (Tamura et al. 2013), the tree was built using the neighbor-joining method (Saitou and Nei 1987). The bootstrap analysis (Felsenstein 1985) with 1000 replicates were performed to assess the support of the clusters. The analysis involved 48 16S rRNA sequences and all positions containing gaps and missing data were eliminated. ▲Clade I, ●Clade II, ■Clade III, colored symbols representing the different clusters are presented in front of each strain.

with two stand-alone proteins containing a Te and an A domain, respectively (Fig. 2B) (Michel et al. 2007; Gasser et al. 2015). Corresponding clustered genes have been found in the chromosome 2 of 19 and 5 strains belonging to clades I and II, respectively (*B. lata* ASM1294v1, all

*B. cenocepia*, *B. pseudomallei* and *B. thailandensis* strains except *B. thailandensis* MSMB121) (Table 1). Except for *B. thailandensis* 2002721723 for which no malleobactin synthesis genes have been found, pyochelin synthesis clusters are always present together with ornibactin or malleobactin synthesis genes. No relationship between occurrence of the pyochelin cluster and the belonging to the clade I or II can be established.

In the chromosome 1 of *B. ambifaria* AMMD, a second cluster characteristic of a siderophore synthesis pathway has been identified beside the cluster for ornibactin synthesis. This cluster includes a gene for a TonB-dependent receptor, genes to build up a diOH-bz (2,3-dihydroxy benzoic acid) monomer and two *nrps* genes coding for a complex constituted of one stand-alone A domain and a protein with a [Cstarter-A-Te] domain architecture (Fig. 2B). The specificity code for the isolated A domain is diOH-bz monomer with a score of 90%. The specificity code for the second A domain is Lys. The organization of the genes and the predictions associated to both A domains are highly compatible with the cepaciachelin siderophore synthesis. The strain was grown under iron-limited conditions allowing the production of siderophores (Fig. S1A). Mass spectrometry analysis of the supernatant revealed the coproduction of two compounds with masses corresponding to C4-ornibactin ([M + H]+ + Fe = 759; [M + Na+] + Fe = 787) and cepaciachelin ([M + H]+ = 789; [M + Na+] = 511) (Fig. S1B). The fragmentation of the latter compounds revealed the presence of diOH-bz and putrescine residues. The production of both siderophores (ornibactin and cepaciachelin) was totally abolished by addition of iron into the culture broth (not shown).

## Identification of a novel lipopeptide synthetase containing dual C/E domains

A complete cluster corresponding to a lipopeptide synthesis pathway has been found in chromosome 2 in the genomes of 9/15 strains belonging to the *pseudomallei* species (Table 1) (Fig. 3). The NRPS is generally constituted of three proteins containing three, four, and five modules, respectively, except for strains 1106b, 1710a, and NCTC13179 for which the mis-annotated synthetase spans on six, four, and five proteins, respectively. The NRPSs are generally annotated as "hypothetical protein," "(non ribosomal) peptide synthetase," or "syringomycin synthetase." The synthetase composed of 12 [C-A-T] modules is predicted to produce a lipopeptide containing 12 monomers in the peptidic moiety. The first C domain has been identified as a Cstarter and the C domains of modules 2, 4, 6, 8, and 11 harbor the dual C/E signature found in CLP domains of *Pseudomonas*. This putative CLP12 is predicted to be FA_D-Ser_Glu_D-Ser_Arg_D-Thr_Leu_D-Dab_Thr_ Thr_D-Glu_Gly_Val (where FA is for Fatty Acid and Dab for diaminobutyric acid). Structure search query performed in the Norine database revealed no peptide displaying similar pattern, even when lowering the threshold to six monomers on the 12 of the peptidic moiety. We suggest to call this potential lipopeptide burkhomycin and to name the corresponding NRPSs BkmA, BkmB, and BkmC (Fig. 3).

The cluster was only partially found in the genome of the six remaining *B. pseudomallei* strains (MSRH 305, 511, 520, NAU20B-16, 668 and NCTC 13178) and in the genomes of two *B. mallei* strains (ATCC 23344 and NCTC 10229) (Fig. 3).

All these NRPSs are characterized by the presence of a Cstarter, a Te tandem ending the assembly line, the lack of any E domain, and the presence of dual C/E domains. A gene encoding a protein of 927 AA, assumed to be an aminotransferase, is present within all the CLP biosynthetic clusters.

## Lipopeptide production by *B. glumae*, *B. gladioli*, and *B. rhizoxinica*

*Burkholderia glumae* BGR1 and *B. gladioli* BSR3, both have a cluster of genes encoding a NRPS potentially producing a lipoheptapeptide (Table 1). Indeed, this protein can be divided into seven modules, the module 1 includes a Cstarter and modules 2, 3, 4, and 6 contain a dual C/E domain. For *B. gladioli* BSR3, the predicted peptidic moiety is D-Thr_D-Pro_D-Gln_Gly_D-X_Phe_Pro, where X represents a position for which no reliable prediction was obtained. For *B. glumae* BGR1, the same structure was predicted except that a D-Ser is predicted for position 3.

The higher number of NRPS gene clusters was found in *B. rhizoxinica* HKI454. The genome of this strain is constituted of one chromosome and two large plasmids pBRH01 and pBRH02, containing 2949, 784, and 203 genes, respectively. The genome mining using the keyword "nonribosomal" revealed at least 31 potentially interesting genes, 12 located on the chromosome and 19 on the plasmid pBRH01. The NRPS clusters represent 4% of the total genome length. We have further analyzed the 19 proteins with a size larger than 1000 AA and we also considered smaller proteins when they were supposed to be included in cluster also containing larger NRPSs. Finally, 13 clusters were pointed out, 10 of them being harbored by the plasmid pBRH01 (Table S3). All the NRPSs contain at least one dual C/E domain and nine of them start with a Cstarter domain, without any ambiguity. The synthetases are predicted to direct the synthesis of 13 different lipopeptides spanning from two to ten monomers within the peptidic moiety. For clusters 10, 11, and 13, the synthetases contain additional methylation domains. The strain was grown in different media (LB, TSB, TSB-glycerol, nutrient broth, Landy). The production of compounds with surfactant properties was assessed by surface tension measurements. The compounds produced by the strain grown in Landy medium displayed the higher activity lowering the surface tension

**Table 1.** Overview on NRPS identified by genome mining on *Burkholderia* strains.

| Phylogenic group | *Burkholderia* species | Genome size (Mb) | Number of chromosomes | Number of plasmids | Lifestyle | Main siderophore | |
|---|---|---|---|---|---|---|---|
| | | | | | | Ornibactin | Malleobactin |
| Clade I | *B. thailandensis* MSMB121 | 6.73 | 2 | 0 | Saprophytic, human pathogen (rare) | | Chr1 |
| | *B. thailandensis* H0587 | 6.76 | 2 | 0 | Saprophytic, human pathogen (rare) | | Chr1 |
| | *B. thailandensis* E264 | 6.72 | 2 | 0 | Saprophytic, human pathogen (rare) | | Chr1 |
| | *B. thailandensis* E444 | 6.65 | 2 | 0 | Saprophytic, human pathogen (rare) | | Chr1 |
| | *B. thailandensis* 2002721723 | 6.57 | 2 | 0 | Saprophytic, human pathogen (rare) | | |
| | *B. pseudomallei* NCTC 13179 | 7.33 | 2 | 0 | Human pathogen | | Chr1 |
| | *B. pseudomallei* MSHR305 | 7.42 | 2 | 0 | Human pathogen | | Chr1 |
| | *B. pseudomallei* MSHR520 | 7.45 | 2 | 0 | Human pathogen | | Chr1 |
| | *B. pseudomallei* K96243 | 7.25 | 2 | 0 | Human pathogen | | Chr1 |
| | *B. pseudomallei* 1710a | 7.33 | 2 | 0 | Human pathogen | | Chr1 |
| | *B. pseudomallei* 668 | 7.04 | 2 | 0 | Human pathogen | | Chr1 |
| | *B. pseudomallei* 1710b | 7.30 | 2 | 0 | Human pathogen | | Chr1 |
| | *B. pseudomallei* NAU20B-16 | 7.31 | 2 | 0 | Human pathogen | | Chr1 |
| | *B. pseudomallei* 1106b | 7.21 | 2 | 0 | Human pathogen | | Chr1 |
| | *B. pseudomallei* 1106a | 7.08 | 2 | 0 | Human pathogen | | Chr1 |
| | *B. pseudomallei* 1026b | 7.23 | 2 | 0 | Human pathogen | | Chr1 |
| | *B. pseudomallei* MSHR511 | 7.31 | 2 | 0 | Human pathogen | | Chr1 |
| | *B. pseudomallei* BPC006 | 7.15 | 2 | 0 | Human pathogen | | Chr1 |
| | *B. pseudomallei* MSHR146 | 7.31 | 2 | 0 | Human pathogen | | Chr1 |
| | *B. pseudomallei* NCTC 13178 | 7.39 | 2 | 0 | Human pathogen | | Chr1 |
| | *B. mallei* NCTC 10229 | 5.74 | 2 | 0 | Human and animal pathogen | | Chr1 |
| | *B. mallei* SAVP1 | 5.23 | 2 | 0 | Human and animal pathogen | | Chr1 |
| | *B. mallei* NCTC 10247 | 5.84 | 2 | 0 | Human and animal pathogen | | Chr1 |
| | *B. mallei* ATCC 23344 | 5.83 | 2 | 0 | Human and animal pathogen | | Chr1 |
| Clade II | *B. multivorans* ATCC 17616 | 7.01 | 3 | 1 | Human pathogen | Chr1 | |
| | *B. multivorans* ATCC 17617 | 7.01 | 3 | 1 | Human pathogen | Chr1 | |
| | *B.* sp. KJ006 | 6.63 | 3 | 1 | Biocontrol, mutualist | Chr1 | |
| | *B. cepacia* GG4 | 6.47 | 2 | 0 | Bioremediation, human pathogen | Chr1 | |
| | *B. cenocepacia* HI2424 | 7.70 | 3 | 1 | Human pathogen | Chr1 | |
| | *B. cenocepacia* AU 1054 | 7.28 | 3 | 0 | Human pathogen | Chr1 | |
| | *B. cenocepacia* J2315 | 8.05 | 3 | 1 | Human pathogen | Chr1 | |
| | *B. cenocepacia* MC0-3 | 7.97 | 3 | 0 | Plant pathogen, human pathogen | Chr1 | |
| | *B. lata* ASM1294v1 | 8.68 | 3 | 0 | Human pathogen | Chr1 | |
| | *B. ambifaria* AMMD | 7.53 | 3 | 1 | Biocontrol | Chr1 | |
| | *B. ambifaria* MC40-6 | 7.64 | 3 | 1 | Human pathogen | Chr1 | |
| | *B. vietnamiensis* G4 | 8.39 | 3 | 5 | Mutualist, human pathogen | Chr1 | |

| Secondary siderophore | | Lipopeptides | | Antifungal | Unknown products | | | |
| --- | --- | --- | --- | --- | --- | --- | --- | --- |
| Pyochelin | Cepaciachelin | Burkhomycin (CLP12) | Cstart_Te (LP) | Burkholdin | "630-2979" AT-CATECAT | PKS-3100_1300_1500 | PKS-NRPS "6274" | NRPS |
| Chr2 | | | | | Chr1 | Chr2 | | |
| Chr2 | | | | | Chr1 | Chr2 | | |
| Chr2 | | | | | Chr1 | Chr2 | | |
| Chr2 | | | | | Chr1 | Chr2 | | |
| Chr2 | | Chr2 | | | Chr1 | Chr2 | | |
| Chr2 | | Chr2(11) | | | Chr1 | Chr2 | Chr1 | |
| Chr2 | | Chr2(9) | | | Chr1 | Chr2 | Chr1 | |
| Chr2 | | Chr2 | | | Chr1 | Chr2 | Chr1 | |
| Chr2 | | Chr2 | | | Chr1 | Chr2 | Chr1 | |
| Chr2 | | Chr2(5) | | | Chr1 | Chr2 | Chr1 | |
| Chr2 | | Chr2 | | | Chr1 | Chr2 | Chr1 | |
| Chr2 | | Chr2(7) | | | Chr1 | Chr2 | | |
| Chr2 | | Chr2 (S) | | | Chr1 | Chr2 | Chr1 | |
| Chr2 | | Chr2 | | | Chr1 | Chr2 | Chr1 | |
| Chr2 | | Chr2 | | | Chr1 | Chr2 | Chr1 | |
| Chr2 | | Chr2(10) | | | Chr1 | Chr2 | Chr1(S) | |
| Chr2 | | Chr2 | | | Chr1 | Chr2 | Chr1 | |
| Chr2 | | Chr2 | | | Chr1 | Chr2 | Chr1(S) | |
| Chr2 | | Chr2(10) | | | Chr1 | Chr2 | Chr1 | |
| | | Chr2(5) | | | Chr1 | | | |
| | | | | | Chr1 | | | |
| | | Chr2(5) | | | | | | |
| | | | | Chr3 | | | | |
| Chr2 | | | | | | | | |
| Chr2 | | | | | | | | |
| Chr2 (P) | | | | | | | | |
| Chr2 | | | | | | | | |
| Chr2 | | | | | | | | |
| | Chr1 | | | Chr3 | | | | |

**Table 1.** (*Continued*)

| Phylogenic group | *Burkholderia* species | Genome size (Mb) | Number of chromosomes | Number of plasmids | Lifestyle | Main siderophore | |
|---|---|---|---|---|---|---|---|
| | | | | | | Ornibactin | Malleobactin |
| Clade III | *B. glumae* BGR1 | 7.28 | 2 | 4 | Plant pathogen | | |
| | *B. gladioli* BSR3 | 9.05 | 2 | 4 | Plant pathogen | | |
| | *B. xenovorans* LB400 | 9.73 | 3 | 0 | Mutualist | | Chr2 |
| | *B.* sp. CCGE1003 | 7.04 | 2 | 0 | Mutualist | *Chr1* | |
| | *B.* sp. CCGE1001 | 6.83 | 2 | 0 | Mutualist | *Chr1* | |
| | *B. rhizoxinica* HKI 454 | 3.75 | 1 | 2 | Mutualist | | |
| | *B.* sp. CCGE1002 | 7.88 | 3 | 1 | Mutualist | | |
| | *B. phymatum* STM815 | 8.68 | 2 | 2 | Mutualist, nitrogen fixation | *Chr2* | |
| | *B.* sp. YI23 | 8.89 | 3 | 3 | Bioremediator | Plasmid | |
| | *B. phytofirmans* PsJN | 8.21 | 2 | 1 | Biocontrol | | Chr2 |
| | *B.* sp. RPE64 | 6.96 | 3 | 2 | Mutualist | Plasmid | |
| | *B. phenoliruptrix* BR3459a | 7.65 | 2 | 1 | Mutualist | | |

Potentially new forms are in italics. Chr, chromosome on which the cluster is; *numbers in brackets indicate the number of monomers expected in the product when different from the model; (P), partial cluster; (S), the NRPS is split into more proteins than in the model; LP, lipopeptide.
*white box is for Chr1; light grey box is for Chr2; dark grey box is for Chr3

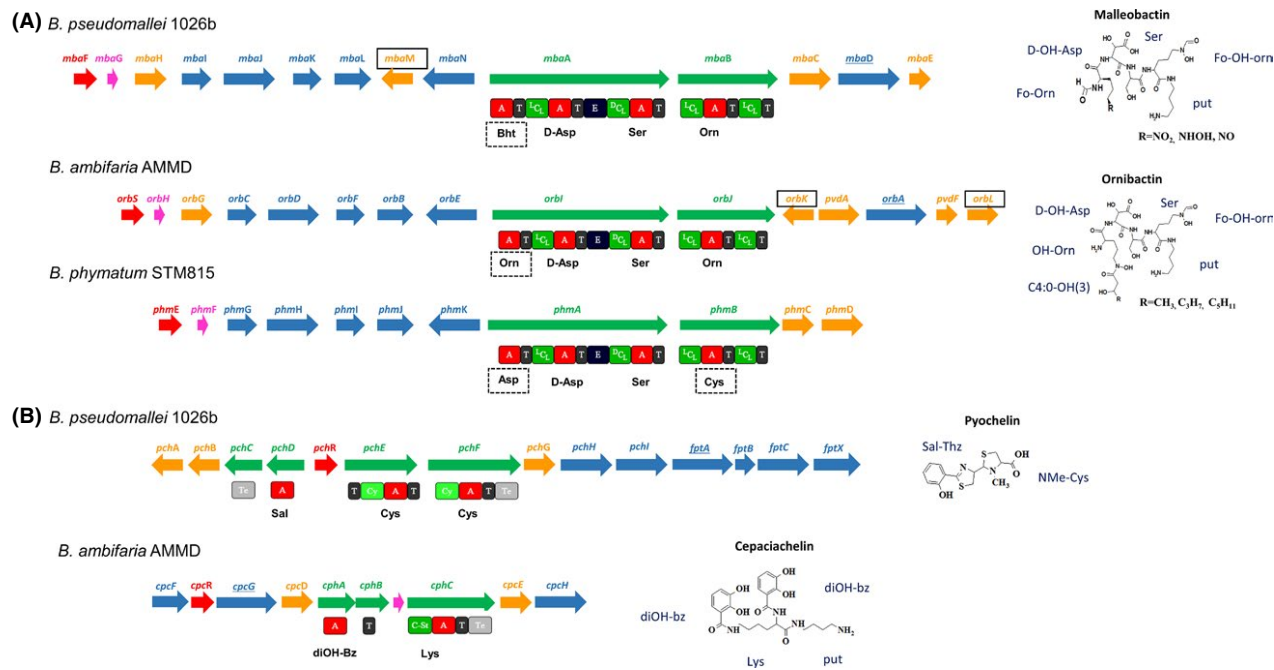

**Figure 2.** Siderophore biosynthesis clusters. Clusters are presented on the left side and the products on the right side. Genes are represented by arrows colored as follows: green for NRPS, orange for accessory enzymes, red for regulatory proteins, blue for uptake participation, and pink for unknown function. Genes encoding TonB-dependent receptor are underlined. The domain organization of the NRPSs is shown below arrows. A, adenylation domain; Cst, C-starter; $^{L}C_{L}$, condensation between two L-monomers; $^{D}C_{L}$, condensation between D-monomer and L-monomer; Cy, heterocyclisation domain; E, epimerization domain; T, thiolation domain; Te, thioesterase domain; predicted amino acid specificity is shown under each A domain; Bht, beta–hydroxy-tyrosin; diOH-bz, 2, 3-dihydroxybenzoic acid; sal, salicylic acid. (A) Main siderophores. Genes and monomers specific for malleobactin or ornibactin synthesis are framed and highlighted by dotted squares, respectively. (B) Secondary siderophores.

| Secondary siderophore | | Lipopeptides | | Antifungal | Unknown products | | | |
|---|---|---|---|---|---|---|---|---|
| Pyochelin | Cepaciachelin | Burkhomycin (CLP12) | Cstart_Te (LP) | Burkholdin | "630-2979" AT-CATECAT | PKS-3100_1300_1500 | PKS-NRPS "6274" | NRPS |
| | | | Chr2 (LP7) | | | | | Chr2 (4a + 4b) |
| | | | Chr2 (LP7 + LP5) | | | | Chr1 | Chr2 (4a + 4b) |
| | | | Chr1 + plasmid (13 LPs) | | | | | |
| | | Chr3(6) | | | | | Chr3 | |

from 70 to 27 mN·m⁻¹. Compared to results obtained with purified surfactin, the curve corresponding to surface tension versus supernatant dilution clearly shows that a mixture of several active compounds is produced by *B. rhizoxinica* HK1454 (Fig. S2). The supernatant was then analyzed by MALDI-ToF. A group of peaks, characteristic of lipopeptide detection with different fatty acid chain lengths, with mass differences of 14 Da (1522 and 1536), was observed. (Fig. S2). The same supernatant was tested for antimicrobial activities. An antibacterial activity was observed against *Micrococcus luteus* and *Listeria innocua*. A moderate inhibition was observed against the yeast *C. albicans* ATCC10231, and no activity at all against fungi *F. oxysporum*, *B. cinerea*, and *R solani* (not shown).

## Identification of the antifungal Bk cluster in the genomes of *B. ambifaria* AMMD and *B.* sp.KJ006

The genome mining revealed an *nrps*-containing cluster located in chromosome 3 of 2 strains: *B. ambifaria* AMMD and *B. sp.* KJ006. This large cluster spanning over 49 000 nt also includes genes coding for regulatory, decorating, and transporter proteins. The proteins exhibit several typical features of NRPS/PKS hybrids involved in Bk biosynthesis (Fig. 4). Eight modules are present, containing NRPS domains as C, A, E, T, and PKS domains as acetytransferase (AT), ACP, ketosynthetase (KS), and ketoreductase (KR) domains. The predicted compound can have the following structure Ile_PKM_Ser_D-Trp_D-Ser_Gly_Asn_D-Ser where the PKM represent a monomer incorporated by the PKS domains. This predicted peptide looks like cyclic glycolipopeptides belonging to the bk/occidiofungin family. Different types of condensation domains have been detected: $^{L}C_{L}$, $^{D}C_{L}$ following an E domain, and an hybrid condensation domain (noted "C") allowing the condensation of a PKM to an NRP monomer (Fig. 4). The presence of a gene encoding a glycosyl transferase suggests that the product can be glysosylated.

A large 9530-bp fragment within gene Bamb_6472 of *B. ambifaria* AMMD was deleted using in vivo homologous recombination in the yeast, *S. cerevisiae* InvSc1 leading to AMMD-Δbamb_6472 mutant (Protocole S2).

Both the AMMD wild type and mutant were tested for biological activities potentially related to the Bk production. Compared to the AMMD wild-type strain, the AMMD-Δbamb_6472-deficient mutant failed to inhibit growth of *C. albicans* and *S. cerevisiae*, exhibited a reduced hemolytic activity, and antagonistic effect on *F. oxysporum*, *G. geotrichum*, *B. cinerea*, and *R solani* (Fig. S3). Culture supernatants from wild-type and deletion mutant both grown in LB broth at 37°C were analyzed by MALDI-ToF. Two peaks observed for m/z 1200.5 and 1222.65 corresponding to the [M+H] and [M+Na] forms of a compound with a mass of 1999.5 Da were only detected in the wild-type supernatant (Protocole S2). Moreover, MALDI-ToF mass spectrometry analysis of culture supernatants confirmed that the production of an antifungal compound with a mass of 1999.5 Da was completely abolished in AMMD-Δbamb_6472 mutant. Considering their belonging to the biosynthesis pathway for bk, the genes were named *bks* for Bk synthesis (Fig. 4).

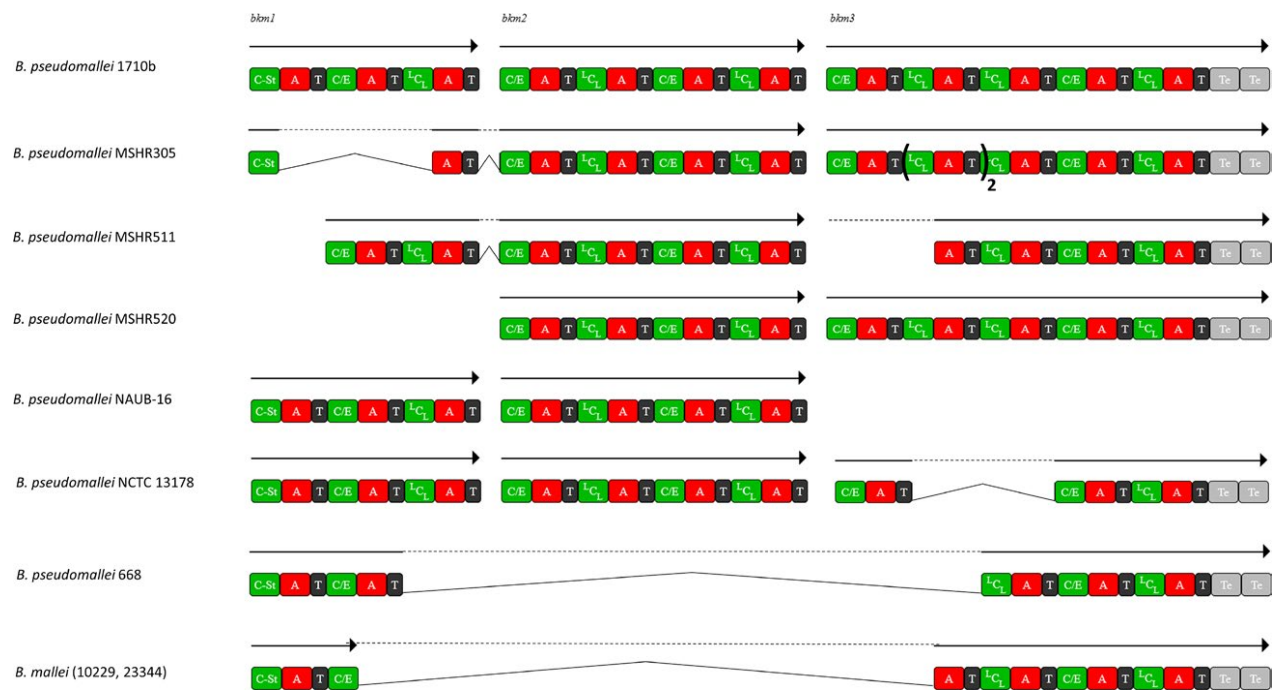

**Figure 3.** Burkhomycin biosynthesis gene clusters. *Burkholderia pseudomallei* 1710b is represented as a model for the complete cluster found in most strains. All the partial clusters are drawn below. Genes are represented by arrows, dotted lines are for lacking genomic sequences, lines represent the fusions in the proteins. Similar domains (adenylation with the same prediction and condensation with the same type) have been aligned. A, adenylation domain; C/E, dual condensation domain catalyzing both epimerization and condensation; LCL, condensation between two L-monomers; Cst, Cstarter for condensation between fatty acid and L-monomer; T, thiolation domain; Te, thioesterase domain.

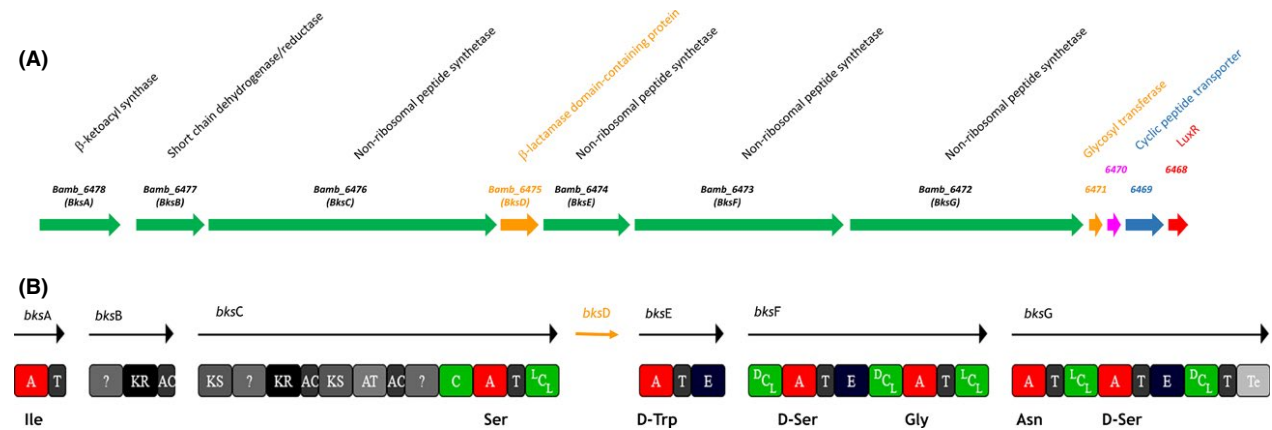

**Figure 4.** Gene cluster responsible for burkholdin (Bk) biosynthesis in *Burkholderia ambifaria* AMMD. (A) all genes involved in Bk synthesis; gene numbers are above the arrows, annotations from NCBI are reported above the gene numbers. Different colors are used to distinguish between the functions: green is for modular enzymes, orange for accessory enzymes, blue for transport, red for regulation, and pink for hypothetical protein. (B) Modular organization of Bk synthesis proteins (Bks). ?, nonidentified domains; A, adenylation domain; C, hybrid condensation domain allowing condensation between a PKS monomer and an NRPS monomer; LCL, condensation between two L-monomers; DCL, condensation between D-monomer and L-monomer; T, thiolation domain; E, epimerization domain; Te, thioesterase domain; AT, acetyltransferase domain; AC, acyl carrier protein domain; KS, ketosynthetase domain; KR, ketoreductase domain. Predicted amino acid specificity is shown under each A domain.

# Discussion

The comprehensive mining carried out on genome sequences has revealed numerous clusters of genes involved in the synthesis of secondary metabolites including genes coding for PKS, NRPS, and hybrids. Among them, more than 160 clusters contain at least one *nrps* gene. With only two exceptions, such *nrps* containing clusters were found in all the explored genomes, whatever phylogenetic distance (Fig. 1) and the bacterial lifestyle (Table 1). Indeed, *Burkholderia* is an intriguing and important genus encompassing a variety of species and strains, ranging from highly pathogenic organisms to strains that promote plant growth (Mahenthiralingam et al. 2008). Some of these properties can be related to the production of secondary metabolites, especially NRPs as those displaying antifungal activities as Bks (Lin et al. 2012). In this article, we have focused on nonribosomal siderophores and lipopeptides because of their potential role in biocontrol, but the annotations we have assigned to some genes will also be very helpful to infer the "*Burkholderia* genome database," a specific tool dedicated to cystis fibrosis research community and aimed to facilitate comparative analysis (Winsor et al. 2008). Elsewhere, the present work completes previous analysis as the annotation of the genome of *B. rhizoxinica* HKI 454 (Lackner et al. 2011). Especially, with this example, we have shown the relevancy of combining approaches both in silico and in wetlab.

The genomic mining for new NRPs highlighted the presence of clusters specific for strains belonging to clade I. This includes a cluster annotated "630-2979" and two hybrid PKS-NRPS "3100-1300-1500" and "6274," named according to the protein sizes (Table 1, Fig. 1). None of the three predicted peptides share any pattern with the known curated peptides annotated in the Norine database. We have no other clues leading to the prediction for siderophore activity or lipopeptide structure and the concerned strains are pathogenic. We can guess that the products may be involved in the virulence and will not be useful in biocontrol, but this has to be further supported experimentally. Moreover, as they seemed to be only produced by bacteria belonging to clade I, they can be considered as maker of the *pseudomallei* species.

As expected, a cluster for biosynthesis of malleobactin and ornibactin was pointed out from the genomes of bacteria belonging to clade I and Bcc group, respectively (Franke et al. 2014, 2015). However, potentially new forms for a main siderophore have been identified in other strains. These includes a malleobactin-like produced by *B. phymatum* we have called phymabactin, and compounds putatively produced by *B. sp.* CCG1001 and *B. sp.* CCG1003. The latter NRPSs contain 6 and 7 modules allowing the production of compounds displaying no resemblance with already known peptides but the clusters are highly suspected to produce a siderophore synthetase because of the presence of genes coding for a TonB-dependent siderophore receptor and an MbtH domain-containing protein (Lautru et al. 2007).

As found in some *Pseudomonas* strains, a pyochelin synthesis cluster has been identified in the genome of strains also producing ornibactin or malleobactin. The cosynthesis of more than one siderophore by bacteria is not well understood, but it has been suggested that beside their ability to provide access to iron, pyochelin, whose affinity for iron is much lower (Cox et al. 1981), may have alternative roles including antibacterial activity (Cornelis and Matthijs 2002; Adler et al. 2012). The pyochelin production may also be associated with the virulence of bacteria (Sokol and Woods 1988). *Burkholderia ambifaria* AMMD does not produce pyochelin, but we have found a cluster of eight genes, located on chromosome 1, directing a siderophore synthesis. The identified cluster includes a gene for a TonB-dependent receptor, genes encoding enzymes necessary to build up a diOH-bz monomer and two *nrps* genes, the predicted monomer for the first one being diOh-bz. All these elements evidenced for the relationship between the biosynthesis cluster and the cepaciachelin known to be produced *B. ambifaria* strain PHP7 (LMG 11351) (Barelmann et al. 1996). Indeed, although the structure of cepaciachelin has been known for a long time (Barelmann et al. 1996), the mechanism of its synthesis was not yet reported. This relationship between genes and the product was supported by MS detection, performed on culture supernatants of the strain grown under iron-limited conditions allowing the production of siderophores (Fig. S1). As the cepciachelin is a nonribosomal peptide, the siderophore has been introduced into the Norine database, the id NOR01254 was attributed.

CLPs are versatile molecules composed of a fatty acid tail linked to a cyclic oligopeptide head, considered as efficient weapons for plant disease biocontrol (Ongena and Jacques 2008). Nearly all CLPs characterized to date have been identified from *Pseudomonas* and *Bacillus* species (Roongsawang et al. 2011). This is comforted with the Norine querying indicating that 139 peptides belonging to 15 families are lipopeptides with surfactant activity. Among them, 76 are produced by *Bacillus* strains and 46 by *Pseudomonas* strains and only 17 are produced by *Serratia*, *Marinobacter*, and *Halomonas*. With the genome mining of *Burkholderia* strains, we have found clusters of genes coding for NRPSs starting with Cstarter and ending with a tandem of Te domains, sharing these features with NRPS of *Pseudomonas* CLPs (De Bruijn et al. 2007). Furthermore, those NRPSs were lacking E domains, and some of the C domains were identified as dual C/E domains. These dual C/E domains were described for the first time in the NRPS responsible for arthrofactin production in *Pseudomonas* (Balibar et al. 2005). Since then, they have been detected in many NRPSs catalyzing the synthesis of CLPs in *Pseudomonas* (Roongsawang et al. 2011) and more recently in *Xanthomonas* (Royer et al. 2013). In all these cases, two types of NRPS clusters coexist in the same strain. The first one contains E domains followed by $^{D}C_L$, generally directing the synthesis of a siderophore (i.e., pyoverdin), while the second type of NRPS containing dual C/E domains is involved in CLP synthesis. Strains can even synthesize several CLPs, with dual C/E domains for all the NRPSs (Pauwelyn et al. 2013; D'aes et al. 2014). Thus, *B. pseudomallei* strains can produce malleobactin where an Asp residue is in D-configuration due to an E domain, and burkhomycin CLP, where five monomers can be epimerized by dual C/E domains. It is quite different with *B. glumae*, *B. gladioli*, and *B. rhizoxinica* because no NRP harboring siderophore activity has been identified. The co-occurrence of two distinct types of NRPs (with E domains vs. with dual C/E domains) in strains sharing the same soil habitat has to be further studied in terms of evolution, also considering the favored horizontal exchanges during the colonization of same ecological niches.

A total of 13 NRPS gene clusters were found in the plasmids (*Burkholderia* sp. I23, *B.* sp. RPE64, *B. glumae* BGR1, *B.gladioli* BSR3, and *B. rhizoxinica* HKI 454). Up to now, this singular feature had only been described in *B. rhizoxinica* (Lackner et al. 2011) and may be specific to the genus *Burkholderia*.

## Acknowledgments

The work was supported by the University of Lille 1, the INTERREG IV program France-Wallonie-Vlaanderen (Phytobio project), the bioinformatics platform Bilille, and Inria. Q. E. received financial support from Sana'a University (Yemen).

## Conflict of Interest

None declared.

## Supporting Information

Additional supporting information may be found in the online version of this article:

**Figure S1.** Detection of siderophores by CAS assays and MALDI-ToF. Left: *Burkholderia phymatum* STM85, right: *Burkholderia ambifaria* AMMD. (A) CAS assays. The color change from blue to orange indicates the presence of iron-chelating compounds. (B) MALDI-ToF spectra. Peaks corresponding to cepaciachelin and ornibactin are pointed out by arrows.

**Figure S2.** Detection of lipopeptides produced by *Burkholderia rhizoxinica* HKI 454. The strain was grown in Landy medium. (A) Surface tension, (B) MALDI-ToF. A group of peaks were observed, characteristic of lipo-peptide detection with different fatty acid chain lengths, with mass differences of 14 Da (1522 and 1536).

**Figure S3.** Biological activities of *Burkholderia ambifaria* AMMD. (A) antifungal activity. *Burkholderia ambifaria* AMMD was inoculated as a line in front of the fungal target on PDA medium. The fungi are mentioned under the plates. Plates were photographed 5 days after inoculation. (B) Antiyeast activity. The yeasts are mentioned under the plates. 1: AMMD wild type, 2: ΔBamb_6472 mutant, 3: sterile water, 4: ethanol 70%. (C) Hemolytic activity on horse bood (5%) LB medium, 50 μL of supernatant obtained after 48 h of growth were used to fill the wells. 1: AMMD wild type, 2: ΔBamb_6472 mutant.

**Table S1.** Strains, plasmids, and primers used in this study.

**Table S2.** Ornibactin, malleobactin, and phymabactin gene clusters. "+" and "−" mention the relative orientation of the genes. Numbers represent the size of the proteins (in AA). Colors represent the role of the protein in the sidero-phore production: green for NRPS biosynthesis, orange for accessory enzymes, blue for uptake, red for regulation, and yellow for unknown function.

**Table S3.** NRPSs of *Burkholderia rhizoxinica* HKI 454. A, adenylation domain; $^{L}C_{L}$, condensation between two L-monomers; $^{D}C_{L}$, condensation between D-monomer and L-monomer; C/E, dual condensation domain catalyzing both epimerization and condensation; T, thiolation domain; E, epimerization domain; Te, Thioesterase domain; "?", nonidentified domain; LP, lipopeptide.

**Protocol S1.** Construction of a *B. ambifaria* AMMD mutant deleted in burkholdin (Bk) synthesis. The supplementary file describes the construction of a mutant strain deleted in the gene *bks*G (Bamb_6472) to support the relationship between the cluster *bks* and the production of an antifungal compound belonging to Bk family.